\documentclass[10pt]{iopart}
\usepackage{iopams}
\usepackage{graphicx}
\usepackage{cite}

\begin{document}
\ioptwocol
\title[duality rotations in media]{The conditions for the preservation of duality symmetry in a linear medium}
\author{Koen van Kruining$^1$ \& J\"org B. G\"otte$^{1,2}$}
\address{$^1$ Max Planck Institute for the Physics of Complex Systems, N\"othnitzer Stra\ss{}e 38, 01187 Dresden, Germany}
\address{$^2$ School of Physics and Astronomy, University of Glasgow, Glasgow, G12 8QQ, UK}
\ead{koen@pks.mpg.de}
\begin{abstract}
Electric magnetic duality symmetry is well understood in vacuum. For light propagating through a medium this symmetry is typically broken. We investigate under what conditions  duality symmetry is preserved in a linear medium and employ these conditions to generalise the definition of optical helicity to a general linear medium. We will discuss some unique properties duality symmetric media possess, provided they exist, and reformulate Maxwell's equation in such a way that they explicitly show the decoupling of opposite helicities. The feasibility of constructing a duality symmetric medium is discussed.
\end{abstract}
\section{Introduction}
In vacuum, nature treats electric and magnetic fields on equal footing \cite{Calkin65}, as a consequence of a symmetry in Maxwell's equations called electric magnetic duality symmetry. For light, interchanging electric and magnetic fields is equivalent to changing one linear polarisation for its orthogonal counterpart. As a consequence, both linear polarisations of light propagate in exactly the same manner as long as electric-magnetic duality symmetry holds. As soon as any piece of material is put in the path of a light beam, this is typically no longer the case. For light impinging on a piece of glass, for instance, the reflection and transmission coefficients are different for different polarisations. Here we investigate what the conditions are for which duality symmetry remains preserved for the passage of light from one medium to another in linear optics. That is, we consider when light can be refracted, reflected and absorbed in a manner independent of its linear polarisation. In order to find the required material properties, we use a continuum description of Maxwell's equations and only consider monochromatic light in order to eliminate dispersion. 

The conserved quantity associated with electric magnetic duality symmetry in vacuum is the optical helicity, which can be interpreted as the amount of linking of the electric and magnetic field lines around each other \cite{TruebaRanada96, AnafasievStepanovski96} or the spin component along the wave vector \cite{CameronBarnettYao,CameronBarnett, BliokhNori11, BliokhBekshaevNori13} or as the difference between the number of left handed and right handed circularly polarised photons \cite{CameronBarnettYao,CameronBarnett, BliokhNori11, BliokhBekshaevNori13, AndrewsColes12}. The topological definition (linking of field lines around each other) is problematic for monochromatic light, since for monochromatic light the field lines do not form closed loops, but go off to infinity, however the other two defintions are unproblematic in vacuum. During the writing of this article, another publication came to our attention which generalises the conservation law of optical helicity to vacuum electrodynamics in the presence of electrical currents, and which can be read as complementary to our work \cite{Nienhuis16}.

The paper is organised as follows. In section \ref{sec:maxwell} we will derive the conditions for duality symmetry in a medium. We then use these conditions to extend the definition of optical helicity to a general linear medium. These are our main results. We will consider some simple examples of duality symmetric media to illustrate the consequences of helicity conservation in section \ref{sec:properties}. In section \ref{sec:separation} we will give a representation of Maxwell's equations that explicitly shows the decoupling of opposite helicity waves in a duality symmetric medium. In section \ref{sec:discussion} the feasability of manufactiuring duality symmetric materials and the broader implications of this work are discussed.

We will refrain from setting the vacuum permittivity, $\epsilon_0$ and the vacuum permeability $\mu_0$ equal to one, in order to easily generalise electrodynamics in vacuum to electrodynamics in a medium.

\section{Maxwell's equations and duality transformations}
\label{sec:maxwell}

As long as there are no charges, electric and magnetic fields are interchangable. This means that under the duality transformation
\begin{eqnarray}
\mathbf D\rightarrow \cos(\theta)\mathbf D-\sin(\theta)\sqrt{\frac{\epsilon_0}{\mu_0}}\mathbf B,\nonumber\\
\mathbf B\rightarrow \cos(\theta)\mathbf B+\sin(\theta)\sqrt{\frac{\mu_0}{\epsilon_0}}\mathbf D,
\end{eqnarray} 
Maxwell's equations retain their form, with $\theta$ being an arbitrary angle \cite{Calkin65}. As soon as any piece of matter is introduced, this duality symmetry breaks at the microscopic level because ordinary matter does contain electrically charged particles, but no magnetically charged particles. 

As long as no charges can get separated over distances on the order of half a wavelength or more, a medium can be viewed as electrically neutral and the breakdown of microscopic duality symmetry is of little importance. This requirement is met in many practical cases (even most lasers do not ionise the air they travel through). If we furthermore assume that the polarisation and magnetisation of the medium induced by the light are linear in the electric and magnetic field strength of the light, a simple modification of Maxwell's equations can describe the effect of the medium on the light fields. In this case, duality symmetry remains preserved at the macroscopic level if the modified Maxwell equations remain invariant under a generalised duality transformation. The preservation of duality symmetry has been investigated for stationary, isotropic, achiral media by Fernandez-Corbaton et.~al.~and the following result can be seen as a generalisation of their work \cite{Australie13}. 

In classical electrodynamics, four different fields are used $\mathbf E$, $\mathbf H$, $\mathbf D$ and $\mathbf B$. $\mathbf E$ and $\mathbf H$ are the `free' electric and magnetic fields and $\mathbf D$ and $\mathbf B$ include the polarisation and magnetisation of the medium as well. For a linear medium, $\mathbf D$ and $\mathbf B$ can be obtained from $\mathbf E$ and $\mathbf H$ by a linear transformation
\begin{equation}
\left[\begin{array}{c} 
\mathbf D\\ \mathbf B
\end{array}\right]=\hat R \left[\begin{array}{c} 
\mathbf E\\ \mathbf H
\end{array}\right],
\end{equation}
with $\hat R$ being the $(6\times 6)$ response matrix \cite{BliokhKivsharNori14} (for monochromatic light)
\begin{equation}
\hat R=\left[\begin{array}{c c}
\epsilon(\mathbf x;\omega) & -\rmi G(\mathbf x;\omega)+\gamma(\mathbf x;\omega)\\ \rmi G(\mathbf x;\omega)+\gamma(\mathbf x;\omega) & \mu(\mathbf x;\omega)
\end{array}\right].
\end{equation}
The entries $\epsilon$ and $\mu$ are the permitivitty and permeability of the medium. These may depend on position and may take the form of matrices if the medium is anisotropic. $G$ is the chiral response present in optically active media and $\gamma$ is the magneto electric respone, proposed by Tellegen \cite{Tellegen48} when he hypothesised a new electronic circuit component and later verified to exist \cite{Dzyaloshinskii60, Astrov60, FolenRadoStalder, Rado}. Like $\epsilon$ and $\mu$, $G$ and $\gamma$ may be anisotropic and may depend on position. The only restriction imposed on $\hat R$ is that it is invertible. In vacuum, the response matrix is
\begin{equation}
\hat R_0=\left[\begin{array}{c c}
\epsilon_0\, I & 0\\ 0& \mu_0\, I
\end{array}\right],
\end{equation}
with $I$ a $3\times 3$ unit matrix. Typically $\epsilon$ (and sometimes $G$) are the only entries which deviate significantly from their vacuum values, but here we assume no restrictions on any of the entries for generality. 

The continuum Maxwell equations are usually written using all four fields, but using the response matrix, two can be eliminated. Keeping only $\mathbf D$ and $\mathbf B$, Maxwell's equations become in matrix form
\begin{eqnarray}
\left[\begin{array}{c c c c c c}
\partial_x & \partial_y & \partial_z &0 & 0 & 0 \\ 0 & 0 & 0 & \partial_x & \partial_y & \partial_z
\end{array}\right]\left[\begin{array}{c} 
\mathbf D\\ \mathbf B
\end{array}\right]=0,\nonumber\\
\left[\begin{array}{c c}
\nabla\times & 0\\ 0&\nabla\times
\end{array}\right]\left[\begin{array}{c c}
0 &I\\ -I & 0
\end{array}\right]\hat R^{-1}\left[\begin{array}{c} 
\mathbf D\\ \mathbf B
\end{array}\right]=\frac \rmd{\rmd t}\left[\begin{array}{c} 
\mathbf D\\ \mathbf B
\end{array}\right],
\end{eqnarray}
It is obvious that any linear combination of $\mathbf D$ and $\mathbf B$ obeys the divergence laws, so a good ansatz to make for the generalised duality transformation is
\begin{equation}
\left[\begin{array}{c} 
\mathbf D\\ \mathbf B
\end{array}\right]\rightarrow\left[\begin{array}{c c} 
\cos\theta I & -\alpha^{-1}\sin\theta I\\ \alpha \sin\theta I & \cos\theta I
\end{array}\right]\left[\begin{array}{c} 
\mathbf D\\ \mathbf B
\end{array}\right],
\end{equation}
with the constant $\alpha$ to be determined later. The duality rotation matrix commutes with the time derivative and with the curl operator. Substituting the ansatz for the duality rotation back into Maxwell's equations shows that duality symmetry is preserved if the duality rotation matrix commutes with
\begin{equation}
\left[\begin{array}{c c}
0 &I\\ -I & 0
\end{array}\right]\hat R^{-1}.
\end{equation}
Using the commutator identity $[A,B^{-1}]=B^{-1}[B,A]B^{-1}$ and evaluating the commutator explicitly leads to the following conditions for duality symmetry 
\begin{eqnarray}
\gamma(\mathbf x;\omega)=0,\nonumber\\
\epsilon(\mathbf x;\omega)\propto\mu(\mathbf x;\omega),\\
G(\mathbf x;\omega)=\mbox{free}.\nonumber
\end{eqnarray}
Where the proportionality applies to both the position dependence and the matrix structure. The constant $\alpha$ for the duality transformation is given by
\begin{equation}
\alpha^2I=\epsilon^{-1}\mu
\end{equation}
The commutator identity used to derive duality symmetry does not only work for $\hat R$ being  a matrix, it works for any linear invertible operator. This makes it possible to generalise the above result to media with a nonlocal linear response. In a nonlocal medium, the polarisation and magnetisation at point $\mathbf x$ do not only depend on the fields in $\mathbf x$, but, due to interactions between neighbouring atoms, on the fields in the surroundings of $\mathbf x$ as well. A linear nonlocal medium can be described by replacing the elements of $\hat R$ with integral kernels \cite{LandauLifshitz8}\footnote{The extra phase $\rme^{\rmi\frac{\omega|\mathbf{x-x'}|}c}$ compared to Landau \& Lifshitz appears because they assume the electric field at one point can affect the polarisation elsewhere instantaneously whereas we assume that a field can only influence points in its future light cone. For the discussion of duality symmetry this difference does not matter.}. 
\begin{equation}
\mathcal D_\alpha(\mathbf x;\omega)=\int \rme^{\rmi\frac{\omega|\mathbf{x-x'}|}c}R_{\alpha\beta}(\mathbf{x,x'};\omega)\mathcal F_\beta(\mathbf x';\omega)\rmd V'.
\end{equation}
Here $\mathcal D_\alpha$ is one of the components of the six component $(\mathbf{D,B})$ vector and $\mathcal F_\alpha$ is one of the components of the six component $(\mathbf{E, H})$ vector and the indices $\alpha$ and $\beta$ run from 1 to 6 and we adopt the usual convention of summing over doubly occurring indices. These integral transforms are linear, so as long as they are invertible as well, the same conditions for duality symmetry still apply. The caveat here is to find out if all integral kernels are invertible. If $R_{\alpha\beta}$ only depends on $\mathbf{x-x'}$, as is the case for homogeneous nonlocal media, invertibility can be checked by Fourier transforming. 

For nonstationary media, whose properties vary with time, an approach similar to the one for nonlocal media can be used\cite{LandauLifshitz8}. Working in the time domain instead of in the frequency domain, the elements of the response matrix for stationary media is given by
\begin{equation}
\mathcal D_\alpha(\mathbf x,t)=\int \int_{-\infty}^{t-\frac{|\mathbf{x-x'}|}c} R_{\alpha\beta}(\mathbf{x,x'},t-t')\mathcal F_\beta(\mathbf x',t')\rmd t'\rmd V'.
\end{equation}
Time varying medium properties can be included by allowing for other time dependences than $t-t'$ in the response matrix and the conditions for duality symetry become.
\begin{eqnarray}
\tilde \gamma(\mathbf {x,x'},t,t')=0,\nonumber\\
\tilde \epsilon(\mathbf {x,x'},t,t')\propto\tilde\mu(\mathbf{x,x'},t,t'),\\
\tilde G(\mathbf {x,x'},t,t')=\mbox{free}.\nonumber
\end{eqnarray}
Where each entry is a $3\times 3$ matrix of integral operators. An important difference between the stationary and the nonstationary case is that in the stationary case one can assume to be working with monochromatic light and require duality symmetry to hold only for the frequency of interest whereas in the nonstationary case the frequency of light can change and duality symmetry must apply broadband.
 
The conserved quantity and current associated with duality symmetry are the helicity density and flux. We propose to generalise these quantities in a linear medium in the following way, using the magnetic vector potential $\mathbf  A$ defined via $\nabla\times \mathbf A=\mathbf B$ and the electric vector potential $\mathbf C$ defined via $-\nabla\times\mathbf C=\mathbf D$:  
\begin{eqnarray}
\mathcal H=\frac 12\left(\alpha^{-1}\mathbf{A^*\bdot B}-\alpha \mathbf{C^*\bdot D}\right),\nonumber\\
\boldsymbol{\Phi}_{\mathcal H}=\frac 12\left(\alpha^{-1}\mathbf{E\times A^*}+\alpha \mathbf{H\times C^*}\right).
\end{eqnarray}
The only addition made compared to previous works \cite{TruebaRanada96, AnafasievStepanovski96, CameronBarnettYao, CameronBarnett, BliokhNori11, BliokhBekshaevNori13} are the prefactors $\alpha^{-1}$ and $\alpha$. Substituting these expressions into the continuity equations and using Maxwell's equations yields
\begin{equation}
\frac{\rmd\mathcal H}{\rmd t}+\nabla\bdot \boldsymbol{\Phi}_{\mathcal H}=-2\mathrm{Re}(\alpha^{-1}\mathbf{E^*\bdot B})+2\mathrm{Re}(\alpha\mathbf{H^*\bdot D}).
\end{equation}
Now assuming duality symmetry, $\mathbf B=\mu\mathbf H+\rmi G\mathbf E$ and $\mathbf D=\epsilon\mathbf E-\rmi G\mathbf H$. Assuming Hermiticity of $\hat R$, $G$ is a real symmetric matrix. This gives
\begin{equation}
\frac{\rmd\mathcal H}{\rmd t}+\nabla\bdot \boldsymbol{\Phi}_{\mathcal H}=-2\mathrm{Re}(\alpha^{-1}\mathbf E^*\bdot \rmi G\mathbf E)+2\mathrm{Re}(\alpha\mathbf H^*\bdot-\rmi G\mathbf H)=0.
\end{equation}
This shows that our helicity is indeed locally conserved if the medium is duality symmetric and if the response matrix is Hermitian, that is, if there is no gain or absorption. 
\section{Properties of duality symmetry preserving interfaces and media}
\label{sec:properties}

Duality symmetric media interact in a unique way with light. An achiral duality symmetric optical element can (if it exists) manipulate light without affecting its its polarisation. To illustrate this property, we will give two simple examples: reflection and transmission off a duality symmetric flat interface and the propagation of light through a duality symmetric uniaxial crystal. 

Consider the interface between two achiral media. The Fresnel coefficients are \cite{Hecht}
\begin{eqnarray}
r_\perp=\frac{\sqrt{\frac{\epsilon_i}{\mu_i}}\cos(\theta_i)-\sqrt{\frac{\epsilon_t}{\mu_t}}\cos(\theta_t)}{\sqrt{\frac{\epsilon_i}{\mu_i}}\cos(\theta_i)+\sqrt{\frac{\epsilon_t}{\mu_t}}\cos(\theta_t)},\nonumber\\ 
t_\perp=\frac{2\sqrt{\frac{\epsilon_i}{\mu_i}}\cos(\theta_i)}{\sqrt{\frac{\epsilon_i}{\mu_i}}\cos(\theta_i)+\sqrt{\frac{\epsilon_t}{\mu_t}}\cos(\theta_t)},\nonumber\\
r_\parallel=\frac{\sqrt{\frac{\epsilon_t}{\mu_t}}\cos\theta_i-\sqrt{\frac{\epsilon_i}{\mu_i}}\cos\theta_t}{\sqrt{\frac{\epsilon_i}{\mu_i}}\cos\theta_i+\sqrt{\frac{\epsilon_t}{\mu_t}}\cos\theta_t},\\
 t_\parallel=\frac{2\sqrt{\frac{\epsilon_t}{\mu_t}}\cos\theta_i}{\sqrt{\frac{\epsilon_i}{\mu_i}}\cos\theta_i+\sqrt{\frac{\epsilon_t}{\mu_t}}\cos\theta_t}.\nonumber
\end{eqnarray}
Duality symmetry implies that the Fresnel coefficients are the same for both polarisations on the interface under consideration. Filling in $\epsilon$ and $\mu$ for the media under consideration shows this and also simplifies the Fresnel coeficients:
\begin{equation}
r_{\perp,\parallel}=\frac{\cos\theta_i-\cos\theta_t}{\cos\theta_i+\cos\theta_t},\qquad
t_{\perp,\parallel}=\frac{2\cos\theta_i}{\cos\theta_i+\cos\theta_t}.
\end{equation}
The refractive index appears only implicitly via Snel's law.

These expressions can also be generalised to chiral duality symmetric interfaces. In that case it is useful to consider circularly rather than linearly polarised light. The Fresnel coefficients follow from the following continuity equations \cite{Chern12}
\begin{eqnarray}
\cos \theta_i\hat x +\rmi\hat y=t_{\mathrm R}(\cos\theta_t \hat x+\rmi\hat y)+r_{\mathrm R}(-\cos\theta_i\hat x+\rmi\hat y),\nonumber\\
\cos \theta_i\hat x -\rmi\hat y=t_{\mathrm L}(\cos\theta_t \hat x-\rmi\hat y)+r_{\mathrm L}(-\cos\theta_i\hat x-\rmi\hat y).
\end{eqnarray}
Here, $\hat x$ and $\hat y$ are the respective unit vectors in the $x$ and $y$ direction. From the continuity equations one obtains
\begin{equation}
r_{\mathrm R}=\frac{\cos\theta_i-\cos\theta_t}{\cos\theta_i+\cos\theta_t},\qquad
t_{\mathrm R}=\frac{2\cos\theta_i}{\cos\theta_i+\cos\theta_t}
\end{equation}
and identical expressions for the left handed polarisation, the only difference being that $\theta_t$ assumes a different value for left and right handed polarisations.   It is interesting to note that for normal incidence $r=0$ and both polarisations are transmitted perfectly. This can be understood by noting that due to rotational symmetry the angular momentum component perpendicular to the interface is conserved. Reflection of circularly polarised light at normal incidence flips the spin angular momentum and therefore may not occur.  Alternatively, the perfect transmission can be understood using the optical impedance \cite{Kronig47}. Interpreting $\frac{E_\parallel}{ H_\parallel}$ as an impedance, one obtains the laws for transmission of an electric signal through an electronic circuit, with transmission being perfect if the impedances of both media are equal. By virtue of the duality symmetry condition $\epsilon\propto \mu$ this impedance matching condition is automatically satisfied for light at normal incidence. 

Another interesting situation to study is the propagation of light through a uniaxial duality symmetric crystal. In most uniaxial crystals, light is separated into an ordinary ray, for which the wave vector is the propagation direction, and an extraordinary ray for which the propagation direction is different from the wave vector. In a duality symmetric crystal, both rays propagate in the same direction, which is different from the wave vector. Both rays can be viewed as `equally extraordinary'. This can be shown as follows. Choose your coordinate system such that the optic axis lies along the z-axis and the wave vector of the light ray under consideration in the xz-plane. The permittivity and permeability can now be written as $\epsilon=\alpha^{-1}\,\mathrm{diag}(c^{-1}_\perp,c^{-1}_\perp,c^{-1}_\parallel)$ and $\mu=\alpha\,\mathrm{diag}(c^{-1}_\perp,c^{-1}_\perp,c^{-1}_\parallel)$, where $\alpha$ is chosen such that the common factor among $\epsilon$ and $\mu$ has the dimension of inverse velocity. One can find two sets of plane wave solutions, the first one for which $\mathbf E_1$ points in the y-direction and the second one for which $\mathbf H_2$ points in the y-direction.
\begin{eqnarray}
\mathbf E_1=\alpha^{\frac 12} \rme^{\rmi(\mathbf{k\bdot x}-\omega t)}\left[\begin{array}{c}0\\1\\0\end{array}\right],\nonumber\\
\mathbf H_2=\alpha^{-\frac 12}\rme^{\rmi(\mathbf{k\bdot x}-\omega t)}\left[\begin{array}{c}0\\1\\0\end{array}\right].
\end{eqnarray}
From Maxwell's equations, one has $\mathbf D_j=-\frac 1\omega \mathbf k\times\mathbf H_j$ and $\mathbf B_j=\frac 1\omega \mathbf k\times\mathbf E_j$. Using these expressions and $\epsilon$ and $\mu$, one can compute $\mathbf H_1$ and $\mathbf E_2$.
\begin{eqnarray}
\mathbf H_1=\frac{\alpha^{-\frac 12}}\omega \rme^{\rmi(\mathbf{k\bdot x}-\omega t)}\left[\begin{array}{c}
-k_zc_\perp\\ 0 \\ k_xc_\parallel
\end{array}\right],\nonumber\\
\mathbf E_2=\frac{\alpha^{\frac 12}}\omega \rme^{\rmi(\mathbf{k\bdot x}-\omega t)}\left[\begin{array}{c}
k_zc_\perp\\ 0 \\ -k_xc_\parallel
\end{array}\right].
\end{eqnarray}
Now the light rays propagate along the (time averaged) Poynting vector, which is $\mathbf S_j=\frac 12\mathbf E_j\times\mathbf H_j^*$. Computing the Poynting vector for both polarisations gives
\begin{equation}
\mathbf S_1=\mathbf S_2=\frac 1\omega\left[\begin{array}{c}
k_x c_\parallel\\ 0 \\ k_z c_\perp \end{array}\right].
\end{equation}
So rays of both polarisations propagate in the same direction which differs from the direction of the wave vector, as depicted in Fig. 1.
\begin{figure}
\includegraphics[width=0.5\columnwidth]{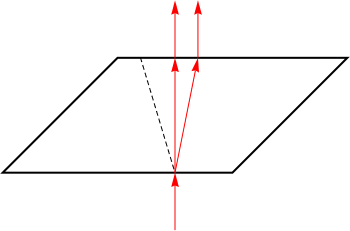}\includegraphics[width=0.5 \columnwidth]{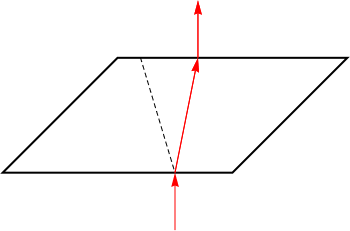}
\caption{Comparison between the passage of rays through an ordinary dielectric uniaxial crystal (left) and a duality symmetric uniaxial crystal (right). In an ordinary uniaxial crystal an incident unpolarised light ray gets separated into two linear orthogonally polarised rays, one of which is shifted relative to the incident ray. In a duality symmetric crystal this separation does not occur and both rays get shifted by the same amount.}
\end{figure}
\section{The separation of the electromagnetic field in a right handed and a left handed field}
\label{sec:separation}

For a duality symmetric medium, there exists another way to write Maxwell's equations in a simple and insightful form, which we will show now. Start by introducing the following right and left handed fields $\mathbf R$ and $\mathbf L$ and their corresponding refractive indices $n_{\mathrm R}$ and $n_{\mathrm L}$:
\begin{eqnarray}
\mathbf R=\frac 1{\sqrt 2}\left(\mathbf E-\rmi\sqrt{\frac\mu\epsilon}\mathbf H\right), \qquad n_{\mathrm R}=\sqrt{\epsilon\mu}+G,\nonumber\\
\mathbf L=\frac 1{\sqrt 2}\left(\mathbf E+\rmi\sqrt{\frac\mu\epsilon}\mathbf H\right), \qquad n_{\mathrm L}=\sqrt{\epsilon\mu}-G.
\end{eqnarray}
When applying these fields and refractive indices in an anisotropic medium, one can remove the ambiguity in the square root of a matrix by replacing $\sqrt{\frac\mu\epsilon}$ by $\alpha^{-1}$ and $\sqrt{\epsilon\mu}$ by either $\alpha\epsilon$ or $\alpha^{-1}\mu$. With these fields the divergence laws take the following form
\begin{eqnarray}
\nabla\bdot \left(\epsilon( \mathbf{R+L})+G\sqrt{\frac\epsilon\mu}(\mathbf{R-L})\right)=0,\nonumber\\
\nabla\bdot\left(\sqrt{\epsilon\mu}(\mathbf{R-L})+G(\mathbf{R+L})\right)=0.
\end{eqnarray}
In a duality symmetric medium, $\sqrt{\frac \epsilon\mu}$ can be taken in front of the divergence operator and the divergence laws can be rewritten as
\begin{equation}
\nabla\bdot(n_{\mathrm R}\mathbf R)=0,\qquad \nabla\bdot(n_{\mathrm L}\mathbf L)=0.
\end{equation}
More interesting are the curl laws, which become, after some algebra,
\begin{eqnarray}
\nabla\times(\mathbf{R+L})=-\rmi\frac \rmd{\rmd t}\left(\sqrt{\epsilon\mu}(\mathbf{R-L})+G(\mathbf{R+L})\right),\\
\nabla\times\sqrt{\frac\epsilon\mu}(\mathbf{R-L})=-\rmi\frac \rmd{\rmd t}\left(\epsilon(\mathbf{R+L})+G\sqrt{\frac \epsilon\mu}(\mathbf{R-L})\right).\nonumber
\end{eqnarray}
Again, if the medium is duality symmetric, a factor of $\sqrt{\frac\epsilon\mu}$ can be taken in front of all differential operators and these equations simplify to
\begin{equation}
\nabla\times\mathbf R=-\rmi\frac \rmd{\rmd t}n_{\mathrm R}\mathbf R,\qquad
\nabla\times\mathbf L=\rmi\frac \rmd{\rmd t}n_{\mathrm L} \mathbf L.
\end{equation}
When written in this form, it is clear that for a duality symmetric medium, right and left handed light fields are completely decoupled and obey their own field equations. Moreover, one can show that any plane wave solution has $\mathbf R$ rotating counterclockwise (with the wave vector pointing toward you) and $\mathbf L$ rotating clockwise, thus verifiying that they indeed contain only resp. right and left handed contributions.

This separation in left and right handed components only works if one takes the fields to be complex. Only then the real part of the electric field and the imaginary part of the magnetic field (and vice versa) interfere destructively for one and constructively for the other handedness. If one treats the field as real quantities $\mathbf L$ becomes the Riemann-Silberstein bivector, which contains all information about both handednesses of the light \cite{Silberstein07a, Silberstein07b, Silberstein12}.  Although Silberstein suggestively used the terms L- and R-quaternions in one of his articles, to the best of our knowledge no one has mentioned this non-intuitive difference between the real and complex treatments of the fields (see also the recent review articles \cite{Birula13, BliokhDresselNori15}). 

Like the fields, the vector potentials can be rewritten in left and right handed contributions as well. Starting out with the magnetic four vector potential $(A_0,\mathbf A)$ and the electric four vector potential $(C_0,\mathbf C)$ with $-\nabla\times \mathbf {C=D}$ and $- \mathbf{\dot C}-\nabla C_0=\mathbf H$, we define left and right handed four vector potentials $(\Lambda_0,\mathbf\Lambda)$ and $(\Delta_0,\mathbf \Delta)$ as
\begin{eqnarray}
(\Lambda_0,\mathbf\Lambda)=\frac 1{\sqrt 2}\left((A_0,\mathbf A)+\rmi\sqrt{\frac\mu\epsilon}(C_0,\mathbf C)\right),\nonumber\\
(\Delta_0,\mathbf \Delta)=\frac 1{\sqrt 2}\left((A_0,\mathbf A)-\rmi\sqrt{\frac\mu\epsilon}(C_0,\mathbf C)\right).
\end{eqnarray}
From these definitions follow
\begin{eqnarray}
\dot\mathbf \Lambda+\nabla\Lambda_0=-\mathbf L,\qquad \nabla\times\mathbf \Lambda=-\rmi n_{\mathrm L}\mathbf L,\nonumber\\
\dot\mathbf\Delta+\nabla\Delta_0=-\mathbf R,\qquad\nabla\times\mathbf \Delta=\rmi n_{\mathrm R}\mathbf R,
\end{eqnarray}
which are similar to the normal relations between vector potentials and fields. It is insightful to rewrite the helicity density and flux in terms of these new fields and potentials. 
\begin{eqnarray}
\mathcal H=\frac\rmi{2\alpha}(\mathbf \Delta^*\bdot n_{\mathrm R}\mathbf R-\mathbf \Lambda^*\bdot n_{\mathrm L}\mathbf L),\nonumber\\
\boldsymbol{\Phi}_{\mathcal H}=\frac1{2\alpha}(\mathbf{R\times\Delta^*+L\times\Lambda^*}).
\end{eqnarray}
These show that the helicity is the difference between left and right handed contributions, which is perfectly in line with the the definition of helicity as the difference in the number of left and right handed circularly polarised photons \cite{CameronBarnettYao, CameronBarnett, BliokhNori11, BliokhBekshaevNori13, AndrewsColes12}. 
\section{Discussion}
\label{sec:discussion}

The condition $\epsilon\propto \mu$ cannot be genercially satisfied for two different materials, because for many materials $\mu$ is close to its vacuum value for most frequencies, whereas $\epsilon$ varies from material to material. Even for materials with strong magnetic responses $\epsilon$ and $\mu$ are not typically proportional across two materials. The reason is that any realistic material is absorbing, which means that the imaginary parts of $\epsilon$ and $\mu$ are nonzero. So there are four parameters\footnote{For isotropic materials; for anisotropic materials the number of parameters is even larger.} which have to be proportional to each other. One of them can be used to set the proportionality constant across the interface, leaving three constraints. If the only thing one can do is varying the frequency of the light, one constraint can be satisfied (provided all parameters vary enough with frequency), leaving two unsatisfied constraints. 

Designing (superconducting \cite{Anlage05, Anlage07}) metamaterials with the desired properties allows to satisfy all constraints within manufacturing tolerance\cite{Pendry95, Pendry98, Pendry99, Schultz00, Schultz01a, Schultz01b}. In this case duality symmetry can be considered an approximate symmetry rather than an exact symmetry.

The only nontrivial case\footnote{That is, other than vacuum or a large piece of homogeneous, isotropic material.} where duality symmetry is exactly satisfied, is a system consisting of a chiral material and the same material with opposite chirality. One can expand the light field in a left handed and a right handed component, and verify that both components experience a change in refractive index as they pass from the right handed to the left handed material or vice versa, yet duality symmetry is obviously preserved here. 

Apart from the above situations, duality symmetry in media is mainly of theoretical interest, because of the central role its associated conserved quantity, helicity, plays in the description of the photon spin.

\ack
We would like to thank Konstantin Bliokh for pointing out the existence of the Tellegen response and Albert Ferrando for useful discussions which led us to separate Maxwell's equations into left and right handed fields. We acknowledge financial support from the UK Engineering and Physical Science Research Council via the grant EP/M01326X/1, the FET grant 295293 of the 7th framework programme of the European Commission and the alumnus programme of the Newton International Fellowship (NF082381) of the Royal Society.

\end{document}